\documentclass[12pt,preprint]{aastex}

\shorttitle{PERIODIC RADIO EMISSION FROM AN M9 DWARF}
\shortauthors{Hallinan et al\.}

\begin{document}


\title{ROTATIONAL MODULATION OF THE RADIO EMISSION FROM THE M9 DWARF TVLM 513-46546: BROADBAND COHERENT EMISSION AT THE SUBSTELLAR BOUNDARY?}


\author{G. Hallinan\altaffilmark{1}, A. Antonova\altaffilmark{2}, J.G. Doyle\altaffilmark{2}, S. Bourke\altaffilmark{1}, W. F. Brisken\altaffilmark{3} and A. Golden\altaffilmark{1}}

\altaffiltext{1}{Computational Astrophysics Laboratory, I.T. Building, National University of Ireland, Galway, Ireland; gregg@it.nuigalway.ie, stephen@it.nuigalway.ie, agolden@it.nuigalway.ie.}

\altaffiltext{2}{Armagh Observatory, College Hill, Armagh BT61 9DG, N. Ireland; tan@arm.ac.uk, jgd@arm.ac.uk.}

\altaffiltext{3}{National Radio Astronomy Observatory, P.O. Box O, Socorro, NM 87801; wbrisken@nrao.edu.}




\begin{abstract}
The Very Large Array was used to observe the ultracool rapidly rotating M9 dwarf TVLM 513-46546 simultaneously at 4.88 GHz and 8.44 GHz.  The radio emission was determined to be persistent, variable and periodic at both frequencies with a period of $\sim 2$ hours.  This periodicity is in excellent agreement with the estimated period of rotation of the dwarf based on its $v$ sin $i$ of $\sim 60$ km s$^{-1}$.  This rotational modulation places strong constraints on the source size of the radio emitting region and hence the brightness temperature of the associated emission.  We find the resulting high brightness temperature, together with the inherent directivity of the rotationally modulated component of the emission, difficult to reconcile with incoherent gyrosynchrotron radiation.  We conclude that a more likely source is coherent, electron cyclotron maser emission from the low density regions above the magnetic poles.  This model requires the magnetic field of TVLM 513-46546 to take the form of a large-scale, stable, dipole or multipole with surface field strengths up to at least 3kG.  We discuss a mechanism by which broadband, persistent electron cyclotron maser emission can be sustained in the low density regions of the magnetospheres of ultracool dwarfs.  A second non-varying, unpolarized component of the emission may be due to depolarization of the coherent electron cyclotron maser emission or alternatively, incoherent gyrosynchrotron or synchrotron radiation from a population of electrons trapped in the large-scale magnetic field.

\end{abstract}





\section{INTRODUCTION}

Magnetic activity in cool main sequence stars is usually diagnosed through the detection of coronal X-ray and radio emission, and via the emission from spectral lines such as chromospheric H$\alpha$. In the Sun's case, it is accepted that such magnetic activity is powered by magnetic fields generated by a so-called $\alpha\Omega$ dynamo at the base of the convective layer where differential rotation is strongest \citep{parker75}.  This dynamo mechanism is also thought to operate in a wide spectral range of cool stars from F-type to early M-type dwarfs and is particularly successful in accounting for the well-established rotation-activity relationship observed over this spectral range \citep{pallavicini81, noyes84}. 

\par For low mass stars with spectral type $\gtrsim$ M3, the nature of magnetic activity is less well understood.  Here lies the boundary at which low mass stars become fully convective, no longer possessing the radiative layer necessary to sustain the $\alpha\Omega$ dynamo.  However, M dwarf flare stars, very often with spectral type later than M3, are associated with intense levels of magnetic activity and are also characterized by large magnetic filling factors, with kG magnetic fields covering a large proportion of the stellar disk \citep{saar94, johnskrull96}.  Indeed, the fraction of objects exhibiting emission associated with magnetic activity, particularly high levels of quiescent H$\alpha$ emission, is seen to increase steadily for spectral types $\gtrsim M3$ reaching 70\% by about spectral type M8 \citep{gizis00, west04}.  Recently, magnetic mapping of the stellar surface of an M4 dwarf has directly confirmed the presence of large-scale, axisymmetric, poloidal fields, including a strong dipolar component, despite the absence of surface differential rotation \citep{donati06}.  Clearly an extremely efficient alternative dynamo operates in low mass, fully convective stars.

\par Beyond spectral type $\sim$ M8, the realm of ultracool stars and brown dwarfs, a sharp drop occurs in the levels of quiescent H$\alpha$ and X-ray emission observed, along with a complete breakdown of the activity-rotation relationship, with many fast rotators ($v$ sin $i$ $>$ 15 km s$^{-1}$) showing no evidence of quiescent H$\alpha$ or X-ray emission \citep{neuhauser99, gizis00, mohanty03, fleming03, west04}.  This reduction in activity may be due to the reduced ionization levels in the atmospheres of these cooler dwarfs \citep{mohanty02}.  Magnetic fields become decoupled from the surrounding atmosphere and are no longer able to generate the necessary magnetic stress required for the dissipation of heat to sustain a chromosphere and corona.   However, this does not preclude the efficient operation of a dynamo or the generation of large-scale kG magnetic fields on ultracool dwarfs.  In fact, despite the reduction in the levels of quiescent emission, there have been detections of H$\alpha$ and X-ray flares from a number of ultracool dwarfs \citep{reid99, gizis00, rutledge00, liebert03, fuhrmeister04, stelzer04, rockenfeller06a} including a T dwarf \citep{burgasser00} which point to the presence of strong magnetic fields.  \citet{rockenfeller06b} have recently used multiband photometric monitoring to confirm a periodicity of 3.65 $\pm 0.1$ hours in observations of the M9V dwarf 2M1707+64.  Moreover, these authors associate this periodicity with the rotational period of the dwarf and rule out the presence of dust clouds as a source of this variability. Instead, they attribute this periodicity to the presence of magnetically induced cool spots on the surface of the star, further evidence for the presence of strong surface magnetic fields on ultracool dwarfs.

\par Based on the sharp drop in quiescent H$\alpha$ and X-ray emission and the empirical relationships established between the observed X-ray and radio emission for a wide spectral range of low mass, main sequence stars \citep{gudel93, benz94}, it was expected that any radio emission from ultracool stars and brown dwarfs would be below the sensitivity of the current generation of radio telescopes.  Nonetheless, \citet{berger01} unexpectedly detected persistent, highly variable radio emission from the M9 field brown dwarf LP 944-20 at levels which, based on the lower limit for the quiescent X-ray flux established by the Chandra observation of \citet{rutledge00}, violated the G\"udel-Benz relations by more than 4 orders of magnitude.  Follow up VLA observations by \citet[hereafter B02]{berger02} yielded three further candidates. All detected sources with measured rotation rates were fast rotators, with confirmed \textit{v} sin \textit{i} ranging up to 60 km s$^{-1}$, despite quiescent levels of H$\alpha$ and X-ray emission being absent, greatly suppressed, or not yet observed.  B02 postulated that the rotation-activity relationship may hold for radio emission in ultracool stars and brown dwarfs despite breaking down in other wavelength regimes.  \citet{burgasser05} also conducted a survey of southern late M and L dwarfs using the Australia Telescope Compact Array (ATCA) and confirmed two detected sources.  One of these sources, DENIS 1048-3956, was seen to emit an extremely bright ($\sim 20$ mJy), narrow bandwidth, $100\%$ circularly polarized flare, with the inferred brightness temperature $\gtrsim 10^{13} K$ requiring a coherent emission mechanism, most likely an electron cyclotron maser which would require the presence of kG magnetic fields.  \citet[hereafter B05]{berger05} have also confirmed a periodicity of $\sim$ 3 hours in the radio emission from the L3.5 candidate brown dwarf 2MASS J00361617+1821104 (hereafter 2M 0036+18) in observations separated by three years, citing three possible sources of the periodicity, (1) orbital motion of a companion (2) rotation of 2M 0036+18 or (3) weak periodic flaring.  Finally, a large scale survey of late M, L and T dwarfs has yielded three more confirmed radio detections and highlights the trend that $L_{\mathrm{rad}}/L_{\mathrm{bol}}$ increases with later spectral type, contrary to what is observed at H$\alpha$ and X-ray wavelengths \citep{berger06}.

\par Observations thus far have established that persistent radio emission at the bottom of the main sequence is a reality, despite the sharp drop in activity observed in X-ray and H$\alpha$ emission and confirming the probable existence of large-scale, stable magnetic fields on ultracool stars and brown dwarfs.  In this paper we report on VLA observations of a previously detected M9 dwarf, TVLM 513-46546 (hereafter TVLM 513), undertaken to further investigate the mechanism responsible for this anomalous radio emission.  These observations provide insights into the nature, topology and strength of the associated magnetic fields and more fundamentally, the dynamo mechanism responsible for their generation.

\section{TVLM 513}

\par TVLM 513 is a young disk M9 dwarf located at a distance of 10.5 pc with a surface temperature of $\approx 2200$ K and a bolometric magnitude of $\log L_{\mathrm{bol}}/L_{\odot} \approx -3.65$ \citep{tinney93, tinney95, leggett01}.  The absence of lithium in the spectrum of TVLM 513 requires its mass to be $> 0.6$ M$_{\odot}$ \citep{reid02}.  We further loosely constrain its mass to be $< 0.8$ M$_{\odot}$ for an age $\tau < 10$ Gyr, based on correlation of its determined properties with the evolutionary models put forward by \citet{baraffe98} and \citet{chabrier00}.  This places TVLM 513 right at the substellar boundary, being either a high mass brown dwarf or an older, very low mass star.  It is one of the most rapidly rotating ultracool dwarfs thus far detected with a measured  \textit{v} sin \textit{i} of $ \sim 60$ km s$^{-1}$ \citep{basri01}.  Despite this rapid rotation it displays only weak H$\alpha$ emission, with equivalent width measurements of 1.7-3.5 \AA{} \citep{martin94, reid01, mohanty03}, and there are no reported X-ray detections in the literature.  Although H$\alpha$ activity is suppressed, it has been strongly detected at radio frequencies.  B02 used the VLA to detect persistent, variable emission at 8.46 GHz which included a strongly right circularly polarized ($\approx 65\%$) flare which reached a flux intensity of $\sim 1100 \mu$Jy.  More recently, \citet{osten06} conducted a multi-frequency VLA observation of TVLM 513 at 8.4 GHz, 4.8 GHz and 1.4 GHz, using a strategy that involved time-sharing a single 10 hour observation between the various frequency bands.  TVLM 513 was detected at each frequency band with only marginal confirmation of variability and no detection of flares or strong circular polarization, possibly due to the absence of continuous coverage at any one frequency band.

\section{RADIO OBSERVATIONS}

\par The radio observations of TVLM513 were conducted with the NRAO Very Large Array (VLA) \footnote[1]{The VLA is operated by the National Radio Astronomy Observatory, a facility of the National Science Foundation operated under cooperative agreement by Associated Universities, Inc.}, for a duration of 4.74 hours on 2005 January 13 and 4.74 hours on 2005 January 14, simultaneously in C (4.88 GHz) and X (8.44 GHz) band. This was achieved by splitting the VLA into two subarrays, with 14 and 11 antennas used for C and X band respectively on January 13, and 13 and 12 antennas used for C and X band respectively on January 14.  3C286 was used for flux calibration and 1513+236 was used for phase calibration with a time on source in a single scan of 12 minutes before moving to the phase calibrator for 2 minutes for both bands.  Reduction of the data was carried out using the AIPS software package. After applying the standard procedures of calibrating and editing the data, the background sources were removed from the visibility data. This was achieved by using the task IMAGR to CLEAN the region around each source, and then the task UVSUB to subtract the resulting models of the background sources from the map. The task UVFIX was used to shift the tangent point coordinates of the target source to coincide with the phase centre.  Light curves were generated by plotting the real part of the complex visibilities at the position of TVLM 513 as a function of time.

\section{PROPERTIES OF THE RADIO EMISSION}

\par TVLM 513 was detected as a variable source at both 4.88 GHz and 8.44 GHz (Figure 1).  The average flux received was $405 \pm 18$ $\mu$Jy at 4.88 GHz and $396 \pm 16$ $\mu$Jy at 8.44 GHz, with the corresponding spectral index of $\alpha = 0.0 \pm 0.2$ consistent with observations of earlier spectral type M dwarfs \citep{gudel96}.  There was significant variability around these average flux levels, as highlighted in Figure 1, which plots the total intensity (Stokes I) radio light curves received at each frequency for both observations with a time resolution of 15 minutes.  In order to quantify this variability we assumed a constant source and determined the reduced chi-square values, $\chi^2_r$, for Stokes I radio light curves derived from the data at each frequency with a range of time resolutions from 5-60 minutes.  The resulting $\chi^2_r$ values range from 1.2 - 4.0 for the 4.88 GHz light curves and 1.4 - 4.8 for the 8.44 GHz light curves.  The higher $\chi^2_r$ values are generally derived from the light curves with lower time resolution, due to reduced error in the data points as a result of lower rms noise in the corresponding longer integration images.  Based on the maximum $\chi^2_r$ values attained, the 4.88 GHz data agree at the 4.63\% level with a constant source while the 8.44 GHz data agree at the 2.9\% level with a constant source.   The upper limit ($3 \sigma$) on the net degree of circular polarization is $r_c \leq 13 \%$ at 4.88 GHz and  $r_c \leq 12 \%$ at 8.44 GHz.  However, in higher time resolution light curves there are significant ($> 3 \sigma$) detections of both left and right circularly polarized emission at both frequencies.  Also, as mentioned above, B02 reported on a $\approx 65\%$ right circularly polarized flare at 8.44 GHz.  As we will discuss further below, we postulate that there may be components of both left and right circularly polarized emission from TVLM 513 which may add destructively to give a much reduced net circular polarization.

\par The radio emission from TVLM 513 is also periodic at both frequencies with a period of $\sim 2$ hours.  This is particularly clear in Figure 2 which correlates the Stokes I radio flux received at 4.88 GHz on January 13th and 14th 2005.  To investigate this periodicity we performed a Lomb-Scargle periodogram analysis \citep{lomb76, scargle82, press89} of the 4.88 GHz and 8.44 GHz Stokes I radio light curves for a wide range of time resolutions.  The significance of the peaks derived from this analysis was dependent on the choice of time resolution, with the strength of the periodic signal reduced for more coarsely sampled data.  With this in mind, the highest time resolution available of 10 seconds is used for both the 4.88 GHz (Figure 3) and 8.44 GHz (Figure 4) periodograms. Two clear peaks are present in both periodograms at frequencies of $\sim$ 0.5 hr$^{-1}$ and $\sim$ 1 hr$^{-1}$ with the peak at 0.5 hr$^{-1}$ corresponding to the fundamental 2 hour period. The secondary harmonic peak at 1 hr$^{-1}$ is accounted for by the presence of two main peaks in the emission per 2 hour period (Figure 1).  The manifestation of both peaks as a group of multiple peaks (spectral leakage) is due to the large gap ($\approx$ 17 hours) in the observation. We used a program provided by Harry Lehto (private communication) which used a 1-dimensional CLEAN algorithm \citep{roberts87} to iteratively remove the spectral window function from the raw data and confirm the presence of both peaks.  To further assert that this periodicity is not an artifact of the observation, we conducted a Lomb-Scargle periodogram analysis of the 4.88 GHz and 8.44 GHz Stokes I light curves of a field source detected in the observation of TVLM 513. A time resolution of 10 seconds was used and the field source data have the same spectral window function as the TVLM 513 data.  No significant peaks were present in the periodogram of the field source data. 

\par In order to determine the significance of the detected periodicity in the radio emission from TVLM 513 we conducted Monte Carlo simulations producing Lomb-Scargle periodograms of 60,000 randomized versions of the data at both frequencies.  For all randomized datasets the same spectral window function as the original data was also maintained.  The resulting false alarm probabilities are shown in Figure 3 and Figure 4.  Both peaks are present with $\gtrsim 99 \%$ probability in the 4.88 GHz data and $> 99.9 \%$ probability in the 8.44 GHz data.  We also include the periodogram of the data from the single observation at 8.44 GHz on January 14 2005 (Figure 5).  Once again, the peaks at $\sim$ 0.5 hr$^{-1}$ and $\sim$ 1 hr$^{-1}$ are present with $> 99.9\%$ probability.  It is also notable that the spectral leakage observed in the periodograms in Figure 3 and Figure 4 is absent in this periodogram which is derived from a single observation and does not contain a large gap in the data.  Interestingly, there are also further significant peaks present in the periodogram at the frequencies of 2 hr$^{-1}$, 3 hr$^{-1}$ and 5 hr$^{-1}$, harmonics of the fundamental 0.5 hr$^{-1}$ frequency.  This may be evidence of further structure in the periodic emission from TVLM 513, in particular the periodic light curve may be non-sinusoidal in nature.  Similar harmonic peaks were not detected in the periodograms of the other individual datasets and further observations are required to investigate this possibility.

\par The data were epoch folded to generate a high signal to noise light curve of the periodic radio emission from TVLM 513.  We initially folded the data for a range of possible periods at or near two hours and analysed the $\chi^2_r$ values of the resulting light curves in an attempt to detect a peak in variance.  A clear peak was observed at approximately 1.96 hours and this value was used as the putative period for epoch folding the data.  Phase bins of 6, 7 and 8 minutes were used with the resulting light curves plotting the total intensity (Stokes I) and circular polarization (Stokes V) of the data shown in Figure 6 (4.88 GHz) and Figure 7 (8.44 GHz).  For each figure two phases of the epoch folded light curve are shown for clarity.  


\par At both frequencies the emission is shown to vary in a non-sinusoidal manner with periodic sharp increases in intensity characterized by high levels of either left or right circular polarization.  At 8.44 GHz there is a short periodic right circularly polarized burst where total intensity levels exceed 1000 $\mu$Jy and circular polarization levels reach $40\%$.  It is  worth noting that a similar burst was observed in a previous 1.83 hour, 8.44 GHz observation of TVLM 513 (B02).  At 4.88 GHz a broad peak in emission observed at phase $\approx 0.3$ is seen to consist of two adjacent but separate peaks, the first showing strong left circular polarization and the second showing right circular polarization.  This amounts to a periodic reversal in the polarization of the emission.  To investigate the variation in circular polarization with frequency we also correlate high time resolution light curves of the Stokes V, 4.88 GHz and 8.44 GHz data, shown in Figure 8.  The large periodic peak in left circular polarization observed in the 4.88 GHz data has no corresponding counterpart in the 8.44 GHz data, possibly indicating a polarization reversal from low to high frequencies.  Another possibility is that different regions producing emission with opposite polarities are contributing to the emission and different regions dominate at different frequencies.  The presence of various regions producing either left or right circularly polarized emission would also account for the periodic reversals in the polarization of the emission at 4.88 GHz.   


\par We can therefore assert that the radio emission from TVLM 513 is characterised by periodic bursts of both left and right circularly polarized emission.  It is noteworthy, however, that the flux intensity does not drop below $\sim 200 \mu$Jy at either frequency and circular polarization is absent for much of the epoch folded light curve.  This indicates that the radio emission from TVLM 513 consists of two components, a highly circularly polarized periodically varying component and a non-varying, unpolarized component.
  
Periodicity was not detected previously by B02, as the corresponding 1.83 hour observation was shorter than the estimated period of $\sim 2$ hours.  Similarly, the observation conducted by \citet{osten06} involved time-sharing between 3 frequency bands with the coarseness of the resulting time-series probably inhibiting the detection of the periodic signal.

\section{SOURCE OF THE PERIODICITY}

\par The most plausible explanation for the periodicity of the radio emission from TVLM 513 is its period of rotation.  The maximum period of rotation can be constrained using the measured \textit{v} sin \textit{i} of $\sim 60$ km s$^{-1}$ and an estimated radius, the latter being a strong function of both mass and age.  Based on a $T_{\mathrm{eff}}$ of 2200 K, the inferred age is greater than 400 Myr, which together with a mass estimate of $0.6$ M$_{\odot} - 0.8$ M$_{\odot}$ yields a radius of $\sim 0.1 \pm 0.01$ R$_{\odot}$ \citep{chabrier00}, which is in agreement with the estimated radius for TVLM 513 of \citet{dahn02}. The resulting maximum period of rotation is $2.0 \pm 0.2$ hours.  This is in excellent agreement with the observed periodicity of the radio emission from TVLM 513 and infers a high inclination angle $i  \gtrsim 65 \arcdeg$, i.e., the equatorial plane is very close to our line of sight.  Considering the striking similarities in the properties of the radio emission from TVLM 513 and the L3.5 dwarf 2M J0036+18 (B02; B05), which include periodicity, a high degree of variability and the periodic presence of high levels of circular polarization, it seems probable that the same emission mechanism is responsible in both cases.  Although, B05 cited 3 possible sources for the 3 hour periodicity in the radio emission from 2M 0036+18, rotation of the dwarf would seem the most likely candidate in light of the probable rotational modulation of the radio emission from TVLM 513.  Based on the \textit{v} sin \textit{i} of $15 \pm 5$ km s$^{-1}$ measured for 2M J0036+18 by \citet{schweitzer01}, this would require a low inclination angle $i \approx 24 \pm 8 \arcdeg$ of the axis of rotation, i.e. almost pole-on.  However, \citet{zapatero06} have recently reported on a revised measurement of $36 \pm 2.7$ km s$^{-1}$, in close agreement with the equatorial velocity derived from a 3 hour period of rotation.  This would require a much larger inclination angle between the axis of rotation and our line of sight $\gtrsim 66 \arcdeg$, similar to what is found for TVLM 513.  
\par Any viable model for the radio emission from these two ultracool dwarfs must account for the rotational modulation, which coupled with the high degree of variability, place strong constraints on the length scale of the emitting regions and therefore the brightness temperature of the associated radio emission.  
 


\section{RADIO EMISSION MECHANISMS} 
 
\par To date, the radio emission from TVLM 513, and other detected ultracool stars and brown dwarfs, has been attributed to incoherent gyrosynchrotron emission from a non-thermal population of mildly relativistic electrons (B02; Burgasser \& Putman 2005; B05; Osten et al. 2006), with the exception of the highly polarized flare detected from DENIS 1048-3956 \citep{burgasser05}.  Such a population of electrons is thought to follow a power law distribution, $n(E) \propto E^{-\delta}$, with $\delta \approx 2-4$ typically invoked for low mass stars \citep{gudel02}.  The resulting radio emission is broadly peaked perpendicular to the field \citep[$\eta_\nu \propto \sin \theta^{-0.43 + 0.65\delta}$,][]{dulk82}, which, if applicable to the periodic component of the radio emission from TVLM 513, requires occultation of emitting active regions by the stellar disc to account for the variability and periodicity.  Indeed, the high degree of variability of the radio emission, characterized by three-fold variations in flux over short timescales $\lesssim$ 0.5 hours (Figure 6 \& Figure 7), requires the distribution of the emitting active regions to be non-uniform and confines their length scale, $L$, to be much less than the size of the stellar disc.  

\par The brightness temperature associated with an incoherent radiation process is limited to $\sim 10^{12}$ K by inverse Compton cooling \citep{kellerman69, readhead94}.  However, this limiting value applies to synchrotron emission from ultra-relativistic electrons in very weak magnetic fields $\ll$ 1 G, as observed from extragalactic sources, whereas here we are interested in gyrosynchrotron emission from mildly relativistic electrons and the corresponding higher strength stellar magnetic fields.  In this case, the emission is emitted at a frequency, $\nu$, where $\nu = s\nu_c$, $s \approx 10 - 100$ and the electron cyclotron frequency $\nu_c \approx 2.8 \times 10^6 B$ Hz. \citet{dulk82} have shown that the effective temperature of gyrosynchrotron emission, $T_{\mathrm{eff}}$, from an isotropic distribution of electrons with a power-law index $2 \lesssim \delta \lesssim 7$ and a low energy cut-off of 10 keV is 

\begin{equation} T_{\mathrm{eff}} \approx 2.2 \times 10^910^{-0.31\delta}(\sin{\theta})^{-0.36-0.06\delta}(\frac{\nu}{\nu_c})^{0.50+0.085\delta} \hspace{0.1cm} K \end{equation} 

\noindent and is limited to a few times $10^9$ K.  The brightness temperature for all incoherent emission is also limited by $T_{\mathrm{B}} \leq T_{\mathrm{eff}}$ \citep{dulk85}, consequently $T_{\mathrm{B}}$ is thus limited to a few times $10^9$ K for gyrosynchrotron emission.

\par The brightness temperature of the radio emission from TVLM 513 is given by

\begin{equation}T_{\mathrm{B}} = 2\times 10^9(f_{\nu}/mJy)(\nu/GHz)^{-2}(d/pc)^2(L/R_{\mathrm{Jup}})^{-2} K \end{equation}  

\noindent Assuming a radius R $\approx 0.1$ R$_{\odot}$ $\sim$ R$_{\mathrm{Jup}}$ infers a $T_{\mathrm{B}} \approx 3.7 \times 10^9 (L/R_{\mathrm{Jup}})^{-2}$, based on the average flux received at 4.88 GHz of 405 $\mu$Jy.  Even assuming a length scale L of the order of the size of the stellar disk would yield a brightness temperature $\sim 1 \times 10^9$ K.  Indeed, this source was also detected at 1.4 GHz by \citet{osten06} with a brightness temperature of $T_{\mathrm{B}} = \sim 2.9 \times 10^{10} (L/R_{\mathrm{Jup}})^{-2}$.  If gyrosynchrotron emission is applicable, such a high brightness temperature requires a source size of the order of the stellar disk or greater.

\par Contrary to this, the characteristics of the periodic light curve require highly directive emission from compact regions much less than the size of the stellar disk with a corresponding much higher brightness temperature.  In particular, the highly circularly polarized bursts at both frequencies (Figure 7 \& Figure 8) have very narrow duty cycles ($\lesssim 0.15$ phase) occurring over timescales much less than that required for an emitting region to rotate in an out of view, thus confirming the directivity of the emission.  Moreover, the periodic reversal of the sense of polarization of the emission at 4.88 GHz confirms the presence of more than one such active region.  It seems unlikely, therefore, that the high brightness temperature and high directivity of the rotationally modulated component of the radio emission from TVLM 513 is compatible with isotropic gyrosynchrotron emission from a large extended corona, although it cannot be ruled out as a possible source of the non-varying, unpolarized component.

\par Other incoherent emission processes such as synchrotron emission from a population of ultra relativistic electrons can be highly beamed perpendicular to the local magnetic field, accounting for the variability and rotational modulation of the radio emission, and can reach higher brightness temperatures than gyrosynchrotron emission.  Indeed, synchrotron radiation has been previously proposed as a source of rotationally modulated stellar radio emission \citep{lim92, lim94}.  However, the detection of significant levels of circular polarization is incompatible with synchrotron emission.  Alternatively, a coherent emission process may be applicable with the possibility of much higher brightness temperatures being invoked.  One such coherent process is plasma radiation, which is generated at the plasma frequency, $\nu_p \approx 9000 n_e^{1/2}$ Hz where $n_e$ is the plasma electron density.  However, plasma radiation is more dominant at lower frequencies ($\lesssim 1$ GHz) due to increasing free-free absorption with increasing $\nu$ (and hence $n_e$) \citep{gudel02}.  It has been speculated that the degree of absorption may be relaxed for a higher temperature, dense, coronal plasma enabling the generation of plasma radiation at higher frequencies \citep{white95, ostenbastian06}.  However, such conditions are at odds with those expected for the increasingly cooler and more neutral atmospheres of ultracool dwarfs as highlighted by the reduced levels of X-ray and H$\alpha$ emission.  

\par Another possible coherent emission process is electron cyclotron maser emission which, as discussed in $\S 7$, can account for the properties of the rotationally modulated emission from both TVLM 513 and 2M 0036+18, particularly if the magnetosphere takes the form of a large-scale, stable, dipole or multipole, with kG field strengths at the surface, such as that confirmed for earlier type M dwarfs by \citet{donati06}.

\section{ELECTRON CYCLOTRON MASER EMISSION}

\par The electron cyclotron maser (hereafter referred to as maser) is a particle-wave plasma instability caused by the resonance between the gyrating electrons in an externally generated magnetic field and the electric field of electromagnetic waves at frequencies near the electron cyclotron frequency and perhaps its low harmonics.  It has been proposed as a possible generation mechanism for a wide range of astrophysical radio emission including planetary radiation from all of the magnetized planets, solar microwave bursts as well as certain emissions from dMe stars, T Tauri stars, RS CVn binaries, Algol-like binaries and magnetic chemically peculiar stars.  

\par To date, masering has been overlooked as a possible source of the radio emission from TVLM 513 and 2M 0036+18 due to the broadband, persistent nature of the detected emission from both candidates \citep[B05,][]{osten06}, whereas maser emission is expected to be confined to short timescales and narrow bands around the fundamental or second harmonic of the electron cyclotron frequency of the emitting source region.  However, the emission is only narrow-banded if the masering is confined to a region with very little variation in the local magnetic field strength and hence, the electron cyclotron frequency.  This may be expected if the anisotropic electron distribution responsible for the maser instability solely takes the form of a loss-cone, a model applied to terrestrial auroral kilometric radiation by \citet{wu79}.  In this model, electrons trapped in a magnetic field mirror at the surface, developing a loss-cone distribution due to the precipitation of electrons with small pitch angles.  This results in an excess in the perpendicular component of the electron velocity distribution perpendicular to the magnetic field $\delta \nu_{\perp}$ such that $\delta f/\delta \nu_{\perp} > 0$, providing the free energy to power the maser. This model was further developed and suggested as a source of certain solar and stellar bursts by \citet{melrose82}, who proposed that mirroring electrons trapped in coronal loops may form a loss-cone at the loop footpoints.  In the case of maser emission powered by a loss-cone, the instability is often confined to regions close to the stellar or planetary surface limiting the range in magnetic field strength and hence the bandwidth of the emission.  It is also subject to saturation of the wave growth as the radiative process removes the loss-cone anisotropy from the electron distribution.  

\par However, work by Pritchett (1984a,b), Pritchett \& Strangeway (1985) and Winglee \& Pritchett (1986) highlighted an alternative mechanism by which the electron distribution can also achieve the necessary anisotropy required for maser emission.  Electrons accelerated along magnetic field lines into regions of higher field strength conserve magnetic momentum by evolving adiabatically to higher pitch angles, generating the necessary excess in the perpendicular component of the electron velocity distribution perpendicular to the magnetic field, $\delta \nu_{\perp}$.  This ``shell distribution" of electrons was subsequently determined to be more dominant in the generation of terrestrial auroral kilometric radiation than the loss-cone distribution \citep[][and references therein]{ergun00}.  Interestingly, maser emission from a shell distribution of electrons is not confined to regions close to the stellar or planetary surface, but can occur from source regions at any height where conditions are suitable for the operation of the maser.  Also, the saturation conditions which limit the growth and continuous generation of maser emission from a loss-cone distribution do not apply to the shell maser, which can sustain steady-state, high brightness temperature emission as long as electron acceleration in the source region is maintained \citep{pritchett99}.  The conditions in the source region must be such that (\textit{a}) the magnetic field strength is relatively high and the plasma electron density resulting in an electron cyclotron frequency $\nu_c$ much greater than the plasma frequency $\nu_p$, where $\nu_c \approx 2.8 \times 10^6$ B Hz and $\nu_p \approx 9000 n_e^{1/2}$ Hz, and (\textit{b}) electrons are continuously accelerated in the source region and can adiabatically evolve to form the shell distribution required to power the maser.  If these conditions were present over a large range of heights above the surface of ultracool dwarfs, with a corresponding wide range of magnetic field strengths, maser emission could viably be continuously generated at a wide range of frequencies resulting in broadband, persistent radio emission.

\par This model, if applicable to ultracool dwarfs, is a significant departure from the model generally applied to cool stars, which attributes the bulk of broadband, persistent radio emission from such stars to incoherent, gyrosynchrotron radiation.  In fact, the coherent radio emission observed from all of the magnetized planets in our solar system \citep[][and references therein]{zarka98}, attributed to the electron cyclotron maser instability, is of more relevance to this discussion.  In particular, we highlight the coherent radio emission from the quasi-dipolar magnetosphere of Jupiter, which originates in the low density, high magnetic latitude regions and from the magnetic flux tube of the volcanic moon Io.  This radio emission is emitted in a number of components at a range of heights above the planetary surface, with frequencies ranging up to a maximum of approximately 40 MHz which corresponds to the electron cyclotron frequency at the regions of strongest magnetic field strength at Jupiter's surface (approx 14 G).  The low cutoff frequency corresponds to the electron cyclotron frequency at the height where $\nu_c/\nu_p$ falls below a critical value, typically between $\sim 2.5$ and 10, thereby quenching the instability \citep{zarka01}. The emission propagates, for the most part, in the supraluminous X mode, resulting in $100\%$ circularly or elliptically polarized emission highly beamed perpendicular to the local magnetic field, that can reach brightness temperatures up to $10^{20}$ K. This X mode emission has an inherent right-handed circular polarization relative to the local magnetic field, resulting in the detection of right circularly polarized emission from Jupiter's northern hemisphere and left circularly emission from Jupiter's southern hemisphere.  Certain components of this radio emission, such as the decametric emission ($\sim 10$ to 40 MHz) from the high magnetic latitudes, referred to as non-Io DAM, can be broadband ($\Delta \nu \sim \nu$), non-varying for timescales of the order of minutes, and show strong rotational modulation.

\par  By comparison, we now consider this maser emission process in the context of the radio emission properties and magnetospheric conditions on ultracool dwarfs such as TVLM 513 and 2M 0036+18.  It has been established that the magnetic fields are large-scale and stable due to the presence of periodicity in radio observations of 2M 0036+18 separated by two years (B05).  If this large scale, stable magnetosphere takes the form of a dipole or multipole, with kG fields present at the surface there may be a large proportion of the magnetosphere where source conditions are suitable for the generation of maser emission.  In particular, considering the atmospheres of dwarf stars become increasingly neutral beyond about spectral type M7 \citep{mohanty02} it is reasonable to assume that the average plasma electron density also drops significantly.  This assumption is supported by the sharp drop in detectable levels of H$\alpha$ and X-ray emission for late M and L dwarfs.  Moreover, if the magnetosphere takes the form of a dipole or multipole, the plasma density would not be homogeneous throughout the magnetosphere.  For example, in the case of a dipole, any plasma injected into the magnetosphere would remain trapped mirroring in the closed field lines and possibly form ``radiation belts" analogous to Earth's Van Allen belts.  However, in regions over the poles, plasma would rapidly escape along open field lines leading to the formation of density cavities in the plasma.  In this scenario, the plasma density would be much higher in the radiation belts than in the ``coronal holes" above the magnetic poles.  The rapid rotation of the dwarf might also further aid the trapping of particles at lower latitudes.  It is feasible that the plasma density in these density cavities is such that $\nu_p \ll \nu_c$ for a wide range of heights above the stellar surface.  Assuming emission at the fundamental of the electron cyclotron frequency, the upper cutoff frequency of emission would probably correspond to the electron cyclotron frequency at the stellar surface while the lower cutoff frequency of emission would correspond to the electron cyclotron frequency at the height above the surface where $\nu_c/\nu_p$ is insufficient and the instability quenches.  In the case of TVLM 513 and 2M 0036+18, with simultaneous detections at frequencies of 4.9 GHz and 8.5 GHz, this would require a range of magnetic field strength in the density cavity ranging from 1.7 - 3 kG with a plasma electron density $\ll 3 \times 10^{11}$ cm$^{-3}$.  As discussed in $\S 1$ the probable detection of an electron cyclotron maser flare at a frequency of 8.5 GHz from the M8.5 dwarf DENIS 1048-3956 \citep{burgasser05} would indicate that kG field strengths may be present on ultracool dwarfs.  A dipolar magnetic field can be modelled by

\begin{equation} B \approx B_s \left( \frac{R_s}{R} \right)^3  \end{equation}

\noindent where $B_s$ and $R_s$ are the surface magnetic field strength and radius respectively.  Considering that the radius of an evolved ultracool dwarf is approximately the same as that of Jupiter, the length scale of the emitting region producing simultaneous emission at 4.9 GHz and 8.5 GHz on TVLM 513 and 2M J0036+18 would be smaller than that producing broadband decametric emission in the frequency range $\sim 10 - 40$ MHz from Jupiter's polar regions, albeit with much stronger kG fields in the source region.  An electron density of $\ll 3 \times 10^{11}$ cm$^{-3}$ in the low density regions above the magnetic poles of an ultracool dwarf is extremely likely, and may be many orders of magnitude less than this value, considering coronal electron density estimates for more massive, hotter, X-ray emitting stars range from $10^{8}$ cm$^{-3} - 10^{13}$ cm$^{-3}$ \citep{gudel04}.  If by some mechanism, a component of the population of trapped mirroring electrons was continuously accelerated into these low density plasma cavities due to the presence of magnetic field-aligned electric fields, these regions could be the source of persistent, broadband, coherent emission.  The generation of such quasi-stable, long-lived, electric fields would have to be inherent to the nature of the large-scale, stable magnetosphere.  One possible source of electric potential drops may be the formation of electrostatic double layers between the populations of trapped ions and electrons.  Ions and electrons mirroring at different locations could possibly form a strong magnetic field-aligned potential drop, which may be a source of the continuous particle acceleration necessary for steady-state maser emission.  

\par The resulting radiation is expected to be highly beamed perpendicular to the magnetic field, explaining the rotational modulation of the emission from both TVLM 513 and 2M J0036+18.  For 2M J0036+18 the emission shows strong net left circularly polarization indicating that the emission from one hemisphere is dominant.  Indeed, Figures 3 \& 4 in B05 would seem to indicate that this circular polarization reaches almost $ 100\%$ once per period of emission.  In the case of TVLM 513, efficient masering from both hemispheres is required to account for the properties of the radio emission, with detections of both left and right circularly polarized emission.  This is unsurprising considering the rotational axis is almost perpendicular to our line of sight.  The presence of two strong peaks per period of emission separated by $\sim$ 0.5 phase also strongly argues for the presence of a dipolar component to the magnetosphere.  The broad peaks in emission occur when the magnetic axis of the dipolar field is perpendicular to our line of sight, which occurs twice per rotational period.  However, the fact that there is no net circular polarization for much of its rotational phase requires a rather contrived geometrical coincidence if the emission is generated solely in the density cavities associated with a dipolar field.  The presence of a multipolar component to the magnetic field, with the corresponding associated density cavities also producing coherent radio emission, may account for this low net polarization. It is also possible that mode conversion or depolarization of the X mode emission occurs in the density cavity.  Mode conversion of terrestrial auroral kilometric radiation from X mode to R mode in the emitting density cavity is proposed by \citet{ergun00} as an efficient method by which the maser emission can escape without reabsorption at higher harmonics of the emission frequency.  Similarly, mode conversion from X mode to O mode has also been considered as a possible source of the elliptical polarization of the Jovian decametric emission \citep[][and references therein]{zarka98}. 

\par Alternatively, the non-varying unpolarized component may be incoherent gyrosynchrotron or synchrotron radiation. Indeed, if the magnetic field takes the form of a large-scale, stable dipole or multipole, the population of electrons mirroring in closed field lines may contribute to the radio emission.  As discussed above, this population of electrons would have an anisotropic distribution which, for gyrosynchrotron emission, can result in higher intensity emission than that produced by an isotropic population \citep{fleishman03}.  It is also possible that such a population of electrons may be continuously accelerated by radial diffusion to extremely high ultrarelativistic energies conducive to the generation of synchrotron emission. Such emission is generated at decimetric wavelengths from the radiation belts of Jupiter and may be a contributing component to the emission from both TVLM 513 and 2M J0036+18.

\section{DISCUSSION AND CONCLUSIONS}

\par We report on VLA observations showing rotational modulation of the radio emission from the M9 dwarf TVLM 513, which places strong constraints on the source size of the emitting region, and hence brightness temperature of the emission.  The resulting high brightness temperatures along with the implicit directivity of the rotationally modulated component of the radio emission are difficult to reconcile with incoherent gyrosynchrotron radiation from a population of mildly relativistic electrons.  We discuss a new model for the radio emission from ultracool dwarfs based on the assumption that the magnetic field on these objects takes the form of a large scale dipole or multipole powered with kG strength fields at the surface.  In regions over the magnetic poles, plasma can escape along open field lines forming density cavities where the electron cyclotron frequency is much greater than the plasma frequency.  If a component of the population of trapped electrons was continuously accelerated into such density cavities by magnetic field-aligned electric fields, possibly due to the presence of electrostatic double layers, such density cavities may be the source of steady-state, broadband electron cyclotron maser emission.  We also discuss the possible presence of a non-varying, unpolarized component of the radio emission, which may be due to depolarization or mode conversion of the maser emission in the density cavity. Alternatively, this component of the radio emission may be incoherent gyrosynchrotron or synchrotron radiation.  In particular, trapped particles mirroring on closed field lines in the large scale magnetic field can form high density radiation belts where electrons can be accelerated to ultrarelativistic energies emitting synchrotron radiation.

\par Coherent maser emission can account for the presence of high levels of circular polarization and the high brightness temperatures of the radio emission from ultracool dwarfs.  The periodicity and high degree of variability from both TVLM 513 and 2M J0036+18 can be attributed to the inherent directivity of the radio emission coupled with the rapid rotation of the dwarf.   We differentiate between this broadband, persistent maser emission and the extremely bright $100\%$ circularly polarized flare observed from the M8.5 dwarf DENIS 1048-3956 (Burgasser \& Putman 2005), typical of flares observed from earlier type M dwarfs.   Such flares are almost always narrow-banded, $100\%$ circularly polarized and can display structure over very short millisecond timescales.  We speculate that both the broadband, quiescent emission and narrowband, flare emission are generated in the same source regions by the same mechanism, the electron cyclotron maser, albeit with fundamental differences in the nature by which the emission is generated.  For example, the smooth broadband emission may be associated with constant particle acceleration due to quasi-stable electric fields across the source region, whereas the narrowband flaring may be due to more impulsive acceleration events.  This is analogous, once again, to Jupiter's decametric emission, which is characterized by smooth, broadband bursts with $\Delta \nu \sim \nu$ (referred to as L bursts), and more impulsive, narrowbanded bursts with $\Delta \nu \ll \nu$, which can reach much higher brightness temperatures (referred to as S bursts) \citep{zarka98}.

\par The configuration of the magnetosphere may also be a contributing factor to the ubiquitous rapid rotation of ultracool dwarfs.  It is unlikely that such a large-scale stable magnetic field would undergo reconnection events in the increasingly neutral atmospheric plasma of ultracool dwarfs, minimizing momentum loss due to dissipation of magnetic energy, and thus leading to much longer spin down times than observed on earlier type dwarfs. This model for the radio emission from ultracool dwarfs can also account for the fact that $L_{\mathrm{rad}}/L_{\mathrm{bol}}$ is maintained, and even seen to increase for late M and L type dwarfs, despite the sharp drop in magnetic activity detected at other frequencies.  The lower plasma density with later spectral type which inhibits the generation of H$\alpha$ and X-ray emission, leads to conditions where more efficient generation of radio emission by the electron cyclotron maser instability can occur.  

\par The mechanism by which H$\alpha$ and X-ray flares are generated on ultracool dwarfs is still unclear.  \citet{chabrier06} postulate that the magnetic stresses necessary for such flares are generated in the higher temperature conductive layers of the star and rise to the surface, dissipating magnetic energy in the upper atmospheric layers.  Such a model is supported by the observations of \citet{fuhrmeister04} who report on a H$\alpha$ flare from the M8.5 dwarf DENIS 1048-3956, the source of the large coherent radio flare discussed in $\S 1$. They interpret broadening on the blue side of the spectral line during the flare as possible evidence that the flare was associated with a rising cloud of emission ejected from the dwarf.  Such ejections may also be the source of the trapped plasma mirroring in the large-scale magnetic field responsible for the generation of the radio emission.    

\par The generation of coherent emission in density cavities associated with large scale dipolar or multipolar fields should be considered as a possible source of both quiescent and flaring radio emission in a wide range of astrophysical objects, including T Tauri stars \citep{smith03}, RS CVn binaries \citep{osten04}, Algol-type binaries \citep{mutel98} and chemically peculiar stars \citep{trigilio00}.  Moreover, the absence of high levels of circular polarization and narrowband structure does not preclude the possibility that such emission may be coherent.  

\par The recent confirmation of a strong, large-scale axisymmetric field on a rapidly rotating M4 dwarf, with a strong dipolar component \citep{donati06}, highlights the possibility that electron cyclotron maser emission from polar density cavities may be a viable source of radio emission from dMe stars.  Indeed, \citet{bingham01} have suggested maser emission from a shell distribution of electrons as a possible source of the 100\% circularly polarized emission from UV Ceti.  Considering the ubiquity of highly polarized, narrowband, coherent bursts from such objects, it is quite possible that broadband, slowly varying maser emission from the same source regions responsible for the flares may also contribute to the quiescent component of the emission.  Indeed, such objects have much longer periods of rotation than later spectral type dwarfs such as TVLM 513 and 2M J0036+18 resulting in much slower rotational modulation of the highly beamed emission, and hence lower variability.  An important factor in the determination of the emission mechanism in dMe stars has been the direct measurement of the extent of stellar radio coronae, and hence constraint of the brightness temperature of the associated emission, afforded by very long baseline interferometry (VLBI).  Some VLBI observations of single dMe stars have revealed large extended radio coronae up to a few stellar radii in diameter, with brightness temperatures and degrees of circular polarization in line with incoherent emission such as gyrosynchrotron or synchrotron radiation \citep{alef97, benz98, pestalozzi00}.  However, other observations of single dMe stars have revealed unresolved flaring and quiescent emission with brightness temperatures, $T_{\mathrm{B}} > 10^{10}$ K, and very high degrees of circular polarization \citep{benz91, benz95}.  In these cases a coherent emission mechanism such as electron cyclotron maser is the more plausible candidate.  Further VLBI observations may clarify the degree to which both coherent and incoherent mechanisms contribute to the radio emission from dMe stars, and whether the prevalence of an individual mechanism is frequency dependent, activity dependent or related to the magnetospheric conditions on each individual star.

\par Further high sensitivity observations of ultracool dwarfs previously detected at radio frequencies should be undertaken to investigate the presence of periodicity and to confirm if this periodicity is indeed due to the rotation of the dwarf.  Direct detection of the magnetic fields on these objects may also be feasible.  Assuming a gyrosynchrotron source would imply a magnetic field strengths of the order of a few hundred gauss to 1kG, however, the coherent process, requires considerably higher field strengths of several kilo-gauss which could potentially be detectable using the FeH band around 1$\mu$m \citep{reiners06}. Although this technique is still in its infancy, the above authors show that that are sufficient FeH lines available even for the faster rotators whose observation and modelling could have an important bearing on the question of which radio process is at work in these ultracool dwarfs.  

\par Confirmation of the generation of broadband coherent emission from ultracool dwarfs would have immense implications for our understanding of stellar magnetic activity and the nature of the dynamo mechanism responsible for the generation of the magnetic fields driving this activity.  It would also highlight the possibility that a similar mechanism for the generation of coherent radio emission from large-scale dipolar or multipolar fields may be a operating in a wide range of astrophysical objects, ranging from planets to brown dwarfs and cool stars.  Comparison with more exotic phenomena such as coherent pulsar radio emission is also compelling.

\acknowledgments

The authors gratefully acknowledge the support of the HEA funded Cosmogrid project and Enterprise Ireland under the grant award SC/2001/0322. The authors also wish to acknowledge the SFI/HEA Irish Centre for High-End Computing (ICHEC) for the provision of computational facilities and support.  Armagh Observatory is grant-aided by the N. Ireland Dept. of Culture, Arts \& Leisure.  We are very grateful to Harry Lehto for the use of his CLEAN algorithm and code and the informative comments on the results produced and to Jerome Sheahan for helpful discussions on certain aspects of this manuscript.  Finally, we would like to thank the referee, Rachel Osten, for valuable input and suggestions on how to improve this manuscript.






\begin{thebibliography}{}
\bibitem[Alef et al.(1997)]{alef97} Alef W., Benz A.O., G\"udel M. 1997, A\&A 317, 707
\bibitem[Baraffe et al(1998)]{baraffe98} Baraffe, I., Chabrier, G., Allard, F., \& Hauschildt, P. H. 1998, A\&A, 337,
403
\bibitem[Basri (2001)]{basri01} Basri, G. 2001, in ASP Conf. Ser. 223, Cool Stars, Stellar Systems and the
Sun: 11th Cambridge Workshop, ed. R. J. Garc\'ia L\'opez, R. Rebolo, \& M. R. Zapaterio Osorio (San Francisco: ASP), 261
\bibitem[Benz \& Alef (1991)]{benz91} Benz A.O., Alef W., 1991, A\&A 252, L19
\bibitem[Benz \& G\"udel (1994)]{benz94} Benz, A. O., \& G\"udel, M. 1994, A\&A, 285, 621
\bibitem[Benz et al.(1995)]{benz95} Benz, A.O., Alef, W., G\"udel, M., 1995, A\&A 298, 187
\bibitem[Benz et al.(1998)]{benz98} Benz, A.O., Conway J.E., G\"udel M., 1998, A\&A 331, 596
\bibitem[Berger (2002)]{berger02} Berger, E. 2002, \apj, 572, 503 (B02)
\bibitem[Berger et al.(2001)]{berger01} Berger, E., et al. 2001, Nature, 410, 338
\bibitem[Berger et al.(2005)]{berger05} Berger, E., et al. 2005, \apj, 627, 960 (B05)
\bibitem[Berger (2006)]{berger06} Berger, E. 2006, \apj, submitted (astro-ph/0603176)
\bibitem[Bingham et al.(2001)]{bingham01} Bingham, R., Cairns, R. A., \& Kellett, B. J. 2001, A\&A, 370, 1000
\bibitem[Burgasser et al.(2000)]{burgasser00} Burgasser, A. J., Kirkpatrick, J. D., Reid, I. N., Liebert, J., Gizis, J. E., \& Brown, M. E. 2000, \aj, 120, 473
\bibitem[Burgasser \& Putman (2005)]{burgasser05} Burgasser, A. J., \& Putman, M. E. 2005, \apj, 626, 486
\bibitem[Chabrier et al.(2000)]{chabrier00}Chabrier, G., Baraffe, I., Allard, F., \& Hauschildt, P. 2000, ApJ, 542, 464
\bibitem[Chabrier \& K\"uker (2006)]{chabrier06}Chabrier, G., \& K\"uker, M. 2006, A\&A, 446, 1027
\bibitem[Dahn et al.(2002)]{dahn02} Dahn, C. C., et al. 2002, \aj, 124, 1170
\bibitem[Donati et al.(2006)]{donati06} Donati, J.-F., Forveille, T., Cameron, A. C., Barnes, J. R., Delfosse, X., Jardine, M. M., Valenti, J. A. 2006, Science, 311, 633
\bibitem[Dulk (1985)]{dulk85} Dulk, G. A. 1985, ARA\&A, 23, 169
\bibitem[Dulk \& Marsh (1982)]{dulk82} Dulk, G. A., \& Marsh, K. A. 1982, \apj, 259, 350
\bibitem[Ergun et al.(2000)]{ergun00} Ergun, R. E., Carlson, C. W., McFadden, J. P., Delory, G. T., Strangeway, R. J., \& Pritchett, P. L. 2000, ApJ, 538, 456
\bibitem[Fleishman \& Melnikov(2003)]{fleishman03} Fleishman, G. D., \& Melnikov V.F. 2003, ApJ 587, 823
\bibitem[Fleming et al.(2003)]{fleming03} Fleming, T. A., Giampapa, M. S., \& Garza, D. 2003, \apj, 594, 982
\bibitem[Fuhrmeister \& Schmitt (2004)]{fuhrmeister04} Fuhrmeister, B., \& Schmitt, J. H. M. M. 2004, A\&A, 420, 1079
\bibitem[Gizis et al.(2000)]{gizis00} Gizis, J. E., Monet, D. G., Reid, I. N., Kirkpatrick, J. D., Liebert, J., \&
Williams, R. J. 2000, \aj, 120, 1085
\bibitem[G\"udel (2002)]{gudel02} G\"udel, M. 2002, ARA\&A, 40, 217
\bibitem[G\"udel (2004)]{gudel04} G\"udel, M. 2004, A\&A Rev., 12, 71
\bibitem[G\"udel \& Benz (1993)]{gudel93} G\"udel, M., \& Benz, A. O. 1993, \apj, 405, L63
\bibitem[G\"udel \& Benz (1996)]{gudel96} Gu\"del, M., \& Benz, A. O. 1996, in ASP Conf. Ser. 93, Radio Emission from the Stars and the Sun, ed. A. R. Taylor \& J.M. Paredes (San Francisco: ASP), 303
\bibitem[Johns-Krull \& Valenti (1996)]{johnskrull96} Johns-Krull, C. M., \& Valenti, J. A. 1996, \apj, 459, L95
\bibitem[Kellerman \& Pauliny-Toth (1969)]{kellerman69} Kellermann, K. I., \& Pauliny-Toth, I. I. K. 1969, ApJ, 155, L71
\bibitem[Leggett et al.(2001)]{leggett01} Leggett, S. K., Allard, F., Geballe, T. R., Hauschildt, P. H., \& Schweitzer, A. 2001, \apj, 548, 908
\bibitem[Liebert et al.(2003)]{liebert03} Liebert, J., Kirkpatrick, J. D., Cruz, K. L., Reid, I. N., Burgasser, A. J., Tinney, C. G., \& Gizis, J. E. 2003, \aj, 125, 343
\bibitem[Lim et al.(1992)]{lim92} Lim, J., Nelson, G. J., Castro, C., Kilkenny, D., \& van Wyk, F. 1992, ApJ, 388,
L27
\bibitem[Lim et al.(1994)]{lim94} Lim, J., White, S. M., Nelson, G. J., \& Benz, A. O. 1994, ApJ, 430, 322
\bibitem[Lomb (1976)]{lomb76} Lomb, N. R. 1976, Ap\&SS, 39, 447
\bibitem[Mart\'in et al.(1994)]{martin94} Mart\'in, E. L., Rebolo, R., \& Magazzu, A. 1994, ApJ, 436, 262

\bibitem[Mohanty \& Basri (2003)]{mohanty03} Mohanty, S., \& Basri, G. 2003, \apj, 583, 451
\bibitem[Mohanty et al.(2002)]{mohanty02} Mohanty, S., Basri, G., Shu, F., Allard, F., \& Chabrier, G. 2002, \apj, 571, 469
\bibitem[Melrose \& Dulk (1982)]{melrose82} Melrose, D. B., \& Dulk, G. A. 1982, ApJ, 259, 844
\bibitem[Mutel et al.(1998)]{mutel98} Mutel, R. L., Molnar, L. A., Waltman, E. B., \& Ghigo, F. D. 1998, ApJ, 507, 371
\bibitem[Neuh\"aser et al.(1999)]{neuhauser99} Neuha\"user, R., et al. 1999, A\&A, 343, 883
\bibitem[Noyes et al.(1984)]{noyes84} Noyes, R.W., Hartmann, L.W., Baliunas, S. L., Duncan, D. K., \& Vaughan, A. H. 1984, ApJ, 279, 763
\bibitem[Osten et al.(2004)]{osten04} Osten, R. A., et al. 2004, ApJS, 153, 317
\bibitem[Osten \& Bastian(2006)]{ostenbastian06} Osten, R. A., \& Bastian, T. S. 2006, ApJ, 637, 1016
\bibitem[Osten et al.(2006)]{osten06} Osten, R. A., Hawley, S. L., Bastian, T. S., \& Reid, I. N. 2006, ApJ, 637, 518
\bibitem[Pallavicini et al.(1981)]{pallavicini81} Pallavicini, R., Golub, L., Rosner, R., Vaiana, G. S., Ayres, T., \& Linsky, J. L. 1981, ApJ, 248, 279
\bibitem[Parker (1975)]{parker75} Parker, E. N. 1975, ApJ, 198, 205
\bibitem[Pestalozzi et al.(2000)]{pestalozzi00} Pestalozzi, M. R., Benz, A. O., Conway, J. E., \& G\"udel, M. 1999, A\&A, 353, 569
\bibitem[Pizzolato et al.(2004)]{pizzolato04} Pizzolato, N., Maggio, A., Micela, G., Sciortino, S., \& Ventura, P. 2003, A\&A, 397, 147
\bibitem[Press \& Rybicki (1989)]{press89} Press, W. H., \& Rybicki, G. B. 1989, \apj, 338, 277
\bibitem[Pritchett (1984a)]{pritchett84a} Pritchett, P. L. 1984a, Geophys. Res. Lett., 11, 143
\bibitem[Pritchett (1984b)]{pritchett84b} Pritchett, P. L. 1984b, J. Geophys. Res., 89, 8957
\bibitem[Pritchett \& Strangeway (1985)]{pritchett85} Pritchett, P. L., \& Strangeway, R. J. 1985, J. Geophys. Res., 90, 9650
\bibitem[Pritchett et al.(1999)]{pritchett99} Pritchett, P. L., Strangeway, R. J., Carlson, C. W., Ergun, R. E., McFadden, J. P., \& Delory, G. T. 1999, J. Geophys. Res., 104, 10317
\bibitem[Readhead (1994)]{readhead94} Readhead, A. C. S. 1994, \apj, 426, 51
\bibitem[Reid et al.(1999)]{reid99} Reid, I. N., Kirkpatrick, J. D., Gizis, J. E., \& Liebert, J. 1999, \apj, 527, L105
\bibitem[Reid et al.(2001)]{reid01} Reid, I. N., Burgasser, A. J., Cruz, K. L., Kirkpatrick, J. D., \& Gizis, J. E. 2001, \aj, 121, 1710
\bibitem[Reid et al.(2002)]{reid02} Reid, I. N.,Kirkpatrick, J. D., Liebert, J., Gizis, J. E., Dahn, C. C.,\& Monet,D.G. 2002, \aj, 124, 519
\bibitem[Reiners \& Basri (2006)]{reiners06} Reiners, A., \& Basri, G. 2006, \apj, 644, 497
\bibitem[Roberts et al (1987)]{roberts87} Roberts, D. H., Lehar, J., \& Dreher, J. W. 1987, AJ, 93, 968
\bibitem[Rockenfeller et al.(2006a)]{rockenfeller06a} Rockenfeller, B., Bailer-Jones, C. A. L., Mundt, R., Ibrahimov, M. A. 2006a MNRAS, 367, 407
\bibitem[Rockenfeller et al.(2006b)]{rockenfeller06b} Rockenfeller, B., Bailer-Jones, C. A. L., Mundt, R. 2006b, A\&A, 448, 1111
\bibitem[Rutledge et al.(2000)]{rutledge00} Rutledge, R. E., Basri, G., Mart\`in, E. L., \& Bildsten, L. 2000, \apj, 538, L141
\bibitem[Saar (1994)]{saar94} Saar, S. H. 1994, in IAU Symp. 154, Infrared Solar Physics, ed. D. M.
Rabin, J. T. Jefferies, \& C. Lindsey (Dordrecht: Kluwer), 493
\bibitem[Scargle (1982)]{scargle82} Scargle, J. D. 1982, \apj, 263, 835
\bibitem[Schweitzer et al.(2001)]{schweitzer01} Schweitzer, A., Gizis, J. E., Hauschildt, P. H., Allard, F., \& Reid, I. N. 2001, ApJ, 555, 368
\bibitem[Smith et al.(2003)]{smith03} Smith, K., Pestalozzi, M., G\"udel, M., Conway, J., \& Benz, A. O. 2003, A\&A, 406, 957
\bibitem[Stelzer (2004)]{stelzer04} Stelzer, B. 2004, \apj, 615, L153
\bibitem[Tinney et al.(1993)]{tinney93} Tinney, C. G., Mould, J. R., \& Reid, I. N. 1993, AJ, 105, 1045
\bibitem[Tinney et al.(1995)]{tinney95} Tinney, C. G., Reid, I. N.,Gizis, J., \& Mould, J. R. 1995, \aj, 110, 3014
\bibitem[Trigilio et al.(2000)]{trigilio00} Trigilio, C., Leto, P., Leone, F., Umana, G., \& Buemi, C. 2000, A\&A, 362, 281
\bibitem[West et al.(2004)]{west04} West, A. A., et al. 2004, \aj, 128, 426
\bibitem[White \& Franciosini (1995)]{white95} White, S. M., \& Franciosini, E. 1995, ApJ, 444, 342
\bibitem[Winglee \& Pritchett (1986)]{winglee86} Winglee, R. W., \& Pritchett, P. L. 1986, J. Geophys. Res., 91, 13531
\bibitem[Wu \& Lee (1979)]{wu79} Wu, C. S., \& Lee, L. C. 1979, \apj, 230, 621
\bibitem[Zapatero Osorio et al.(2006)]{zapatero06} Zapatero Osorio, M. R., Mart\'in, E. L., Bouy, H., Tata, R., Deshpande, R. \& Wainscoat, R. 2006, \apj, submitted (astro-ph/0603194)
\bibitem[Zarka (1998)]{zarka98} Zarka, P. 1998, J. Geophys. Res., 103, 20159
\bibitem[Zarka et al.(2001)]{zarka01} Zarka, P., Queinnec, J. \& Crary, F. J. 2001, P\&SS, 49, 1137


\end{thebibliography}

\clearpage

\begin{figure}[h]

\plotone{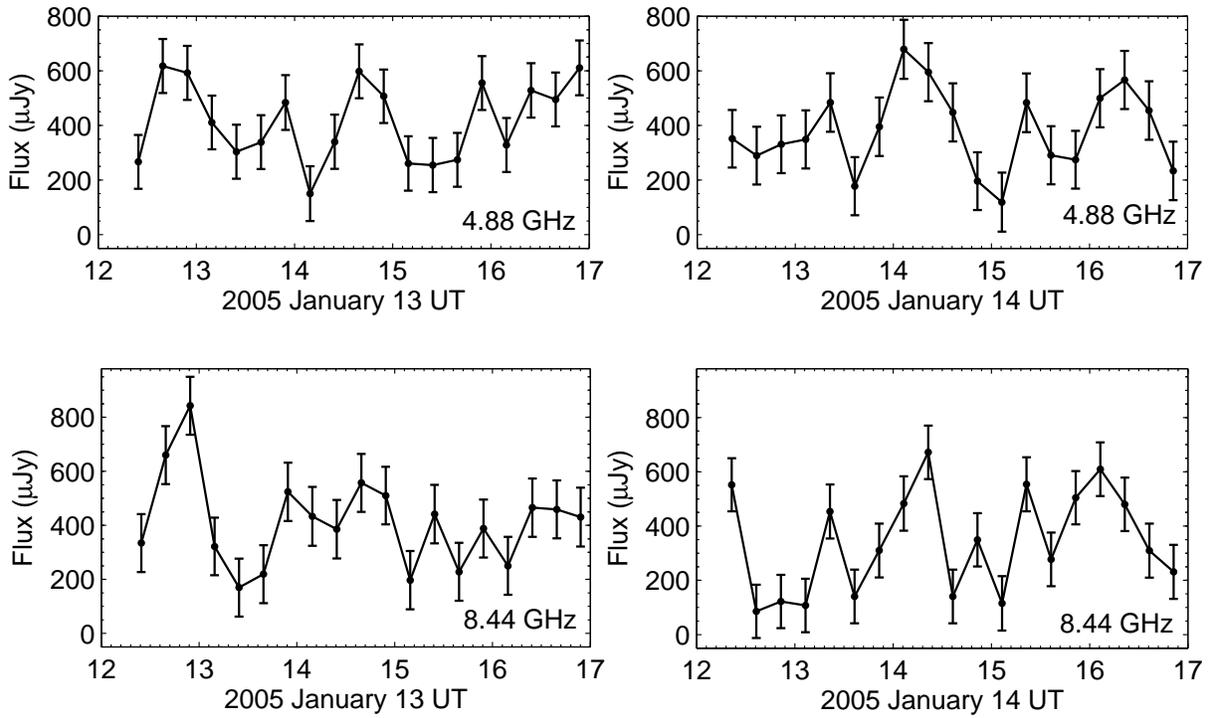}
\caption{Light curves for the Stokes I radio flux received from TVLM 513 at 4.88 GHz (\textit{top}) and 8.44 GHz (\textit{bottom}) on 2005 January 13 and 2005 January 14.  The time resolution used is 15 minutes for all light curves.}

\end{figure}

\begin{figure}[h]
\plotone{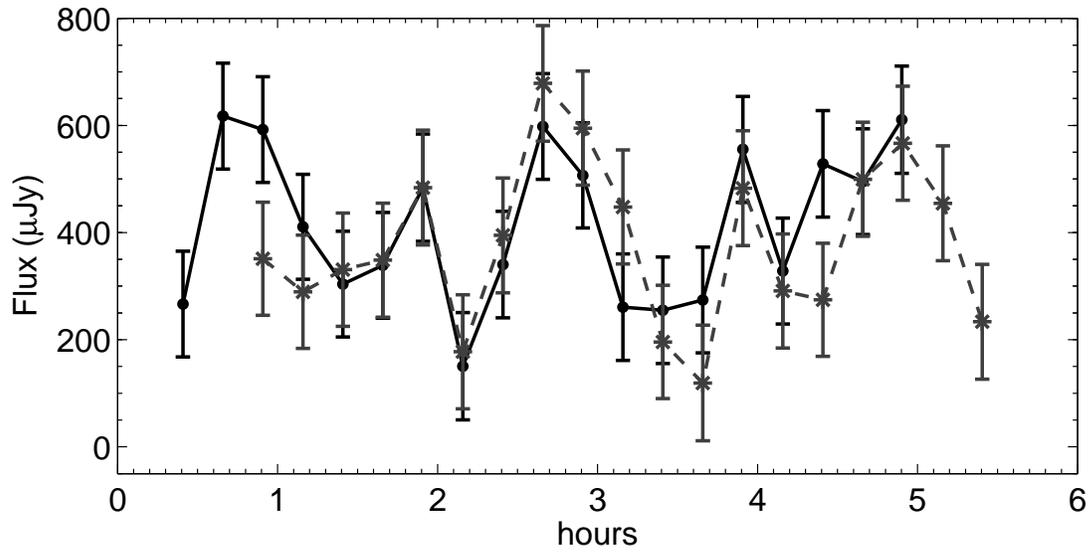}
\caption{Light curves for the Stokes I radio flux received from TVLM 513 at 4.88 GHz on 2005 January 13 (\textit{solid line}) and 2005 January 14 (\textit{dashed-dotted line}). The two light curves have been aligned to highlight the close correlation and periodicity.}     
\end{figure}

\begin{figure}[h]
\plotone{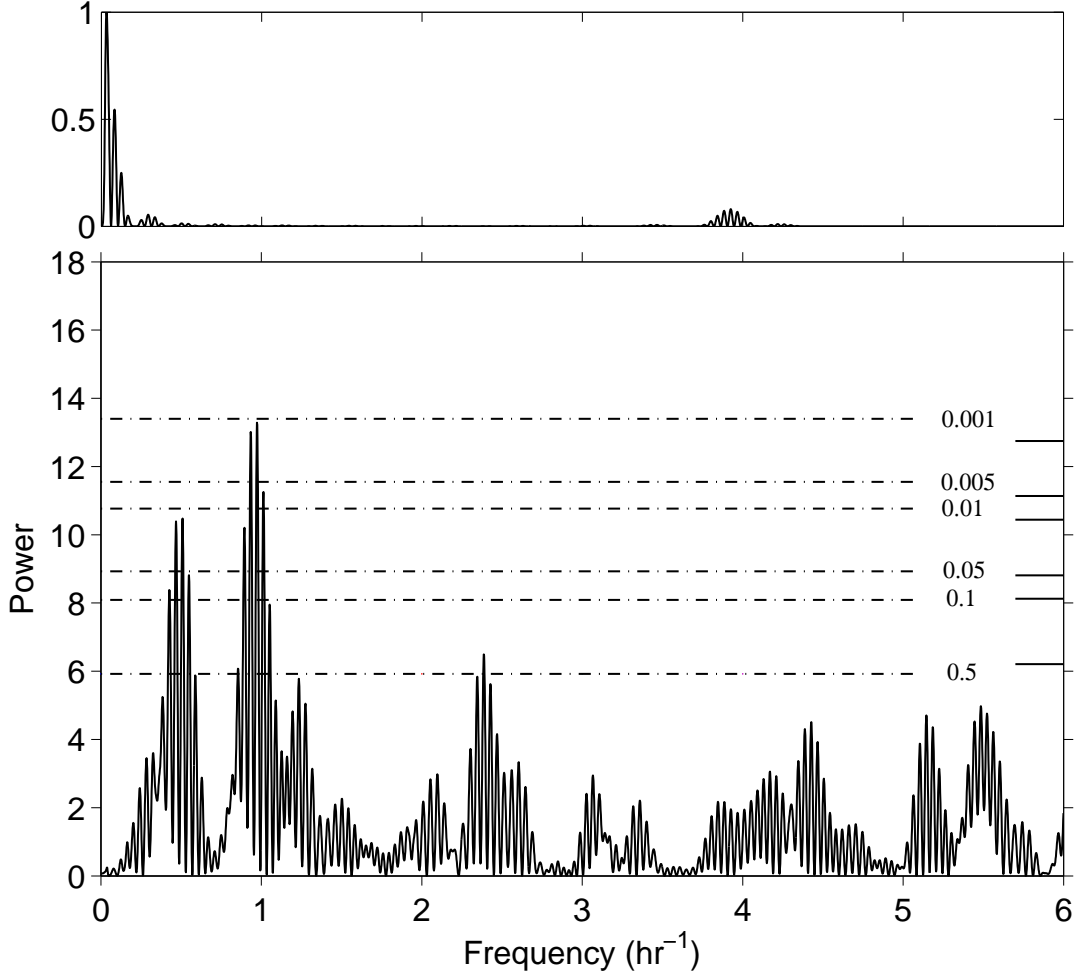}
\caption{Lomb-Scargle periodogram of the 4.88 GHz Stokes I radio emission from TVLM 513 with a time resolution of 10 seconds.  Clear peaks are present at 0.5 hr$^{-1}$ and 1 hr$^{-1}$, corresponding to the two peaks in emission per 2 hour period.  The spectral leakage present for each peak is due to the spectral window function (shown on top) which is a result of the 17 hour gap between the two observations.  The small peak in the spectral window function at $\sim$ 4 hr$^{-1}$ is due to the phase calibration every 15 minutes.   The dashed-dotted lines represent the false alarm probabilities of the significance of the peaks based on 60,000 Monte-Carlo simulations of randomised versions of the light curve with the same spectral window function as the original data.  The solid lines on the right hand side of the graph represent the false alarm probabilities of the significance of the peaks as calculated by the Lomb-Scargle algorithm.}     
\end{figure}

\begin{figure}[h]
\plotone{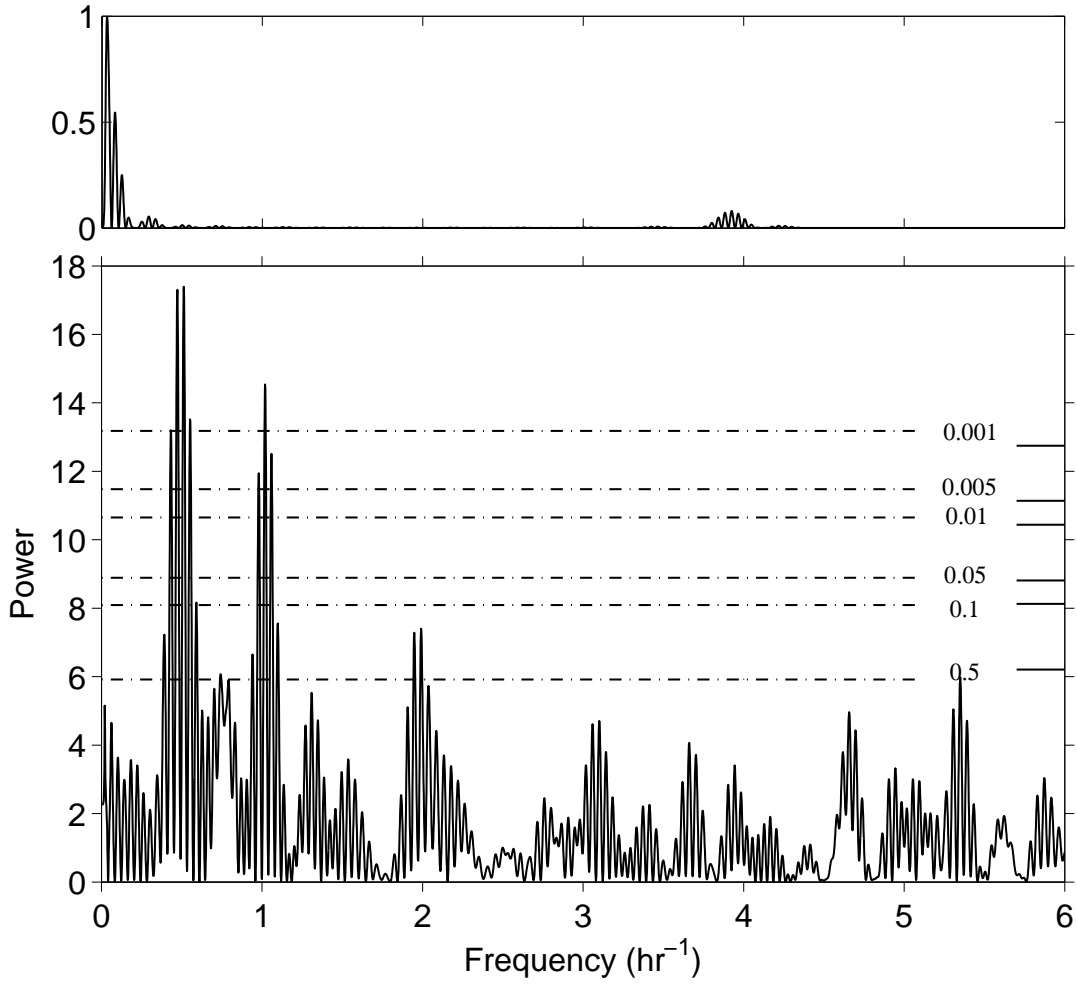}
\caption{Same as Figure 3 for the 8.44 GHz Stokes I radio emission from TVLM 513 with a time resolution of 10 seconds.  Once again, two clear peaks at 0.5 hr$^{-1}$ and 1 hr$^{-1}$ are present.}
\end{figure}

\begin{figure}[h]
\plotone{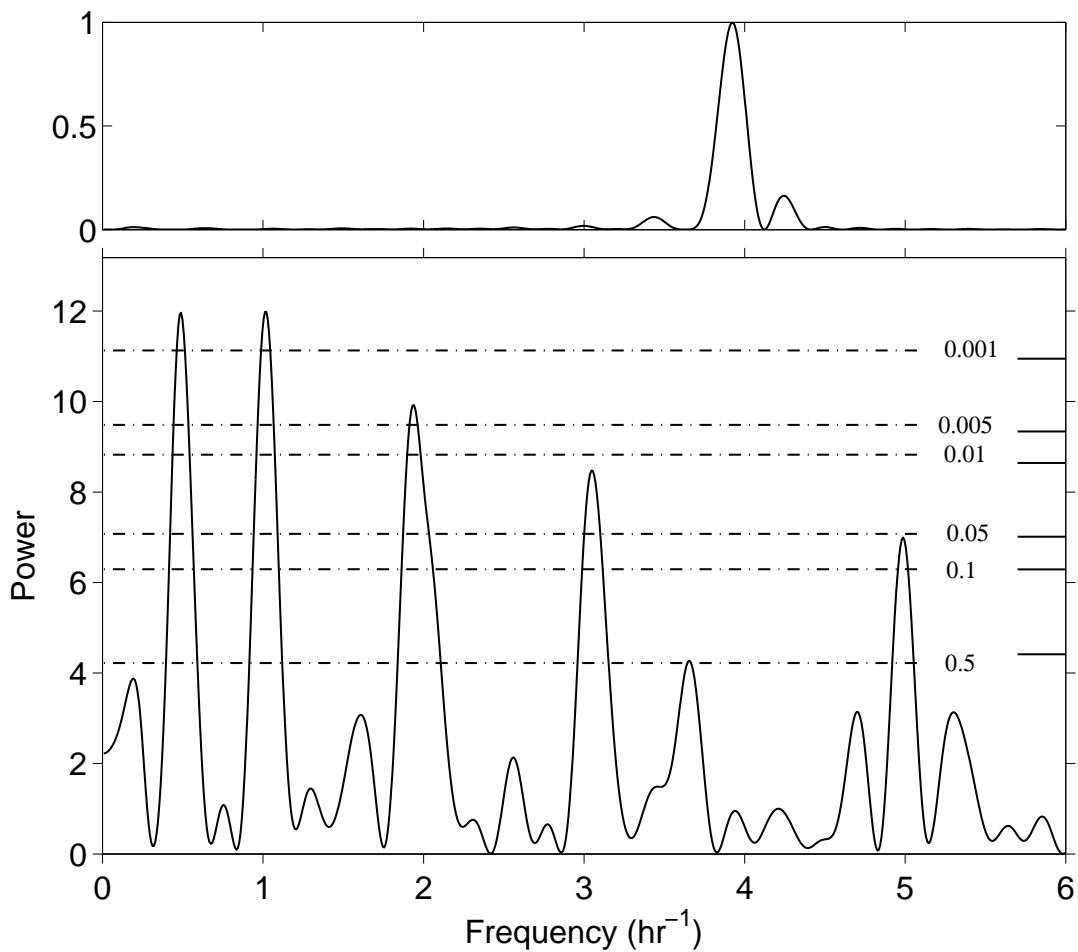}
\caption{Lomb-Scargle periodogram of the 8.44 GHz Stokes I radio emission from the single observation of TVLM 513 on 2005 January 14, with a time resolution of 10 seconds.  There is no spectral leakage due to the absence of a large gap in the data.  The single peak in the spectral window function at $\sim$ 4 hr$^{-1}$ is due to the phase calibration every 15 minutes.  Once again, the two clear peaks at 0.5 hr$^{-1}$ and 1 hr$^{-1}$ are present.   Significant peaks at the frequencies of 2 hr$^{-1}$, 3 hr$^{-1}$ and 5 hr$^{-1}$, harmonics of the fundamental 0.5 hr$^{-1}$ frequency, are also present.  This highlights the possible presence of further structure in the periodic radio light curve of TVLM 513, in particular the periodic light curve may be non-sinusoidal in nature.}     
\end{figure}

\begin{figure}[h]
\plotone{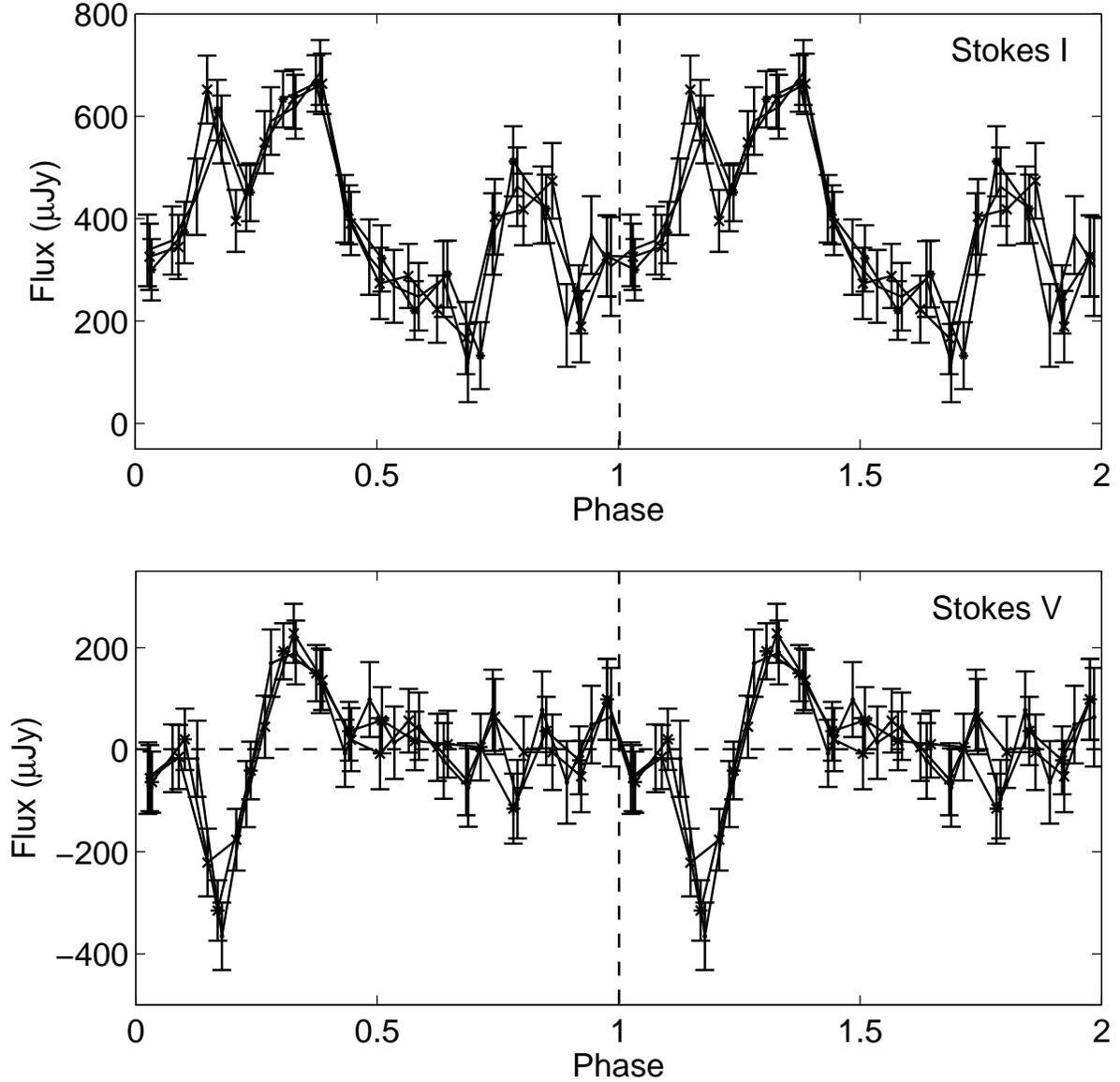}
\caption{Epoch folded light curves derived from the 4.88 GHz Stokes I (\textit{top}) and Stokes V (\textit{bottom}) radio data from TVLM 513 with phase bins of 6, 7 \& 8 minutes.  Two periods of the same epoch folded light curve are shown for clarity.  Positive values for Stokes V indicate right circular polarization while negative values indicate left circular polarization. The main peak at phase 0.3 is resolved to consist of two adjacent peaks, the first one showing strong left circular polarization with the second showing right circular polarization.  These highly polarized emissions add destructively over the course of one period of emission resulting in much lower net circular polarization.} 
\end{figure}

\begin{figure}[h]
\plotone{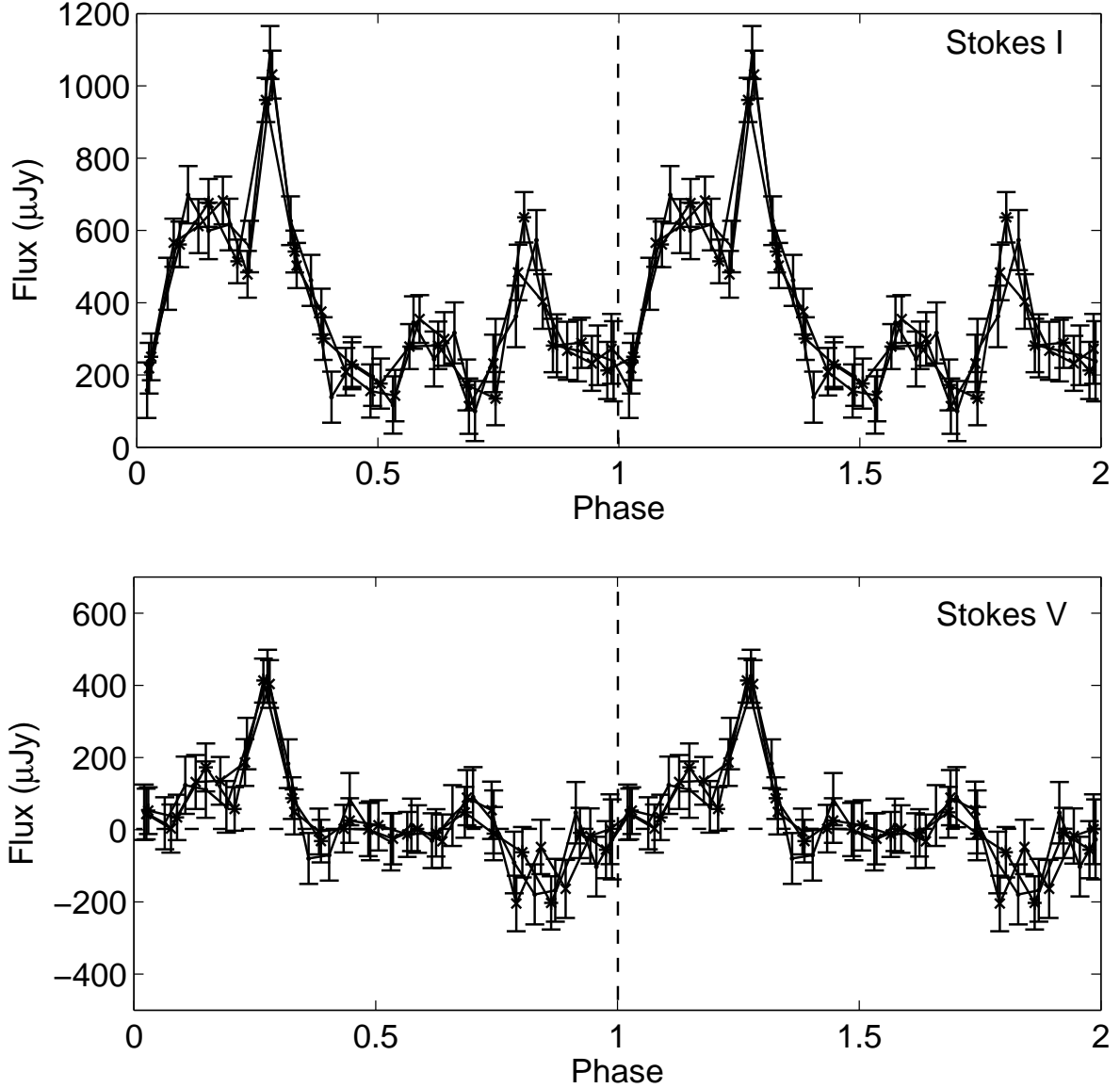}
\caption{Epoch folded light curves derived from the 8.44 GHz Stokes I (\textit{top}) and Stokes V (\textit{bottom}) radio data from TVLM 513 with phase bins of 6, 7 \& 8 minutes.  Two periods of the same epoch folded light curve are shown for clarity.  Positive values for Stokes V indicate right circular polarization while negative values indicate left circular polarization.  The presence of periods of right circular polarization is clear while evidence for left circular polarization is marginal at best.  Of particular interest is the short peak of right circularly polarized emission where Stokes I flux levels exceed $1000 \hspace{0.1cm} \mu$Jy.  The presence of such short duration periodic peaks implies that the emission is inherently directive or beamed in nature.}     
\end{figure}

\begin{figure}[h]
\plotone{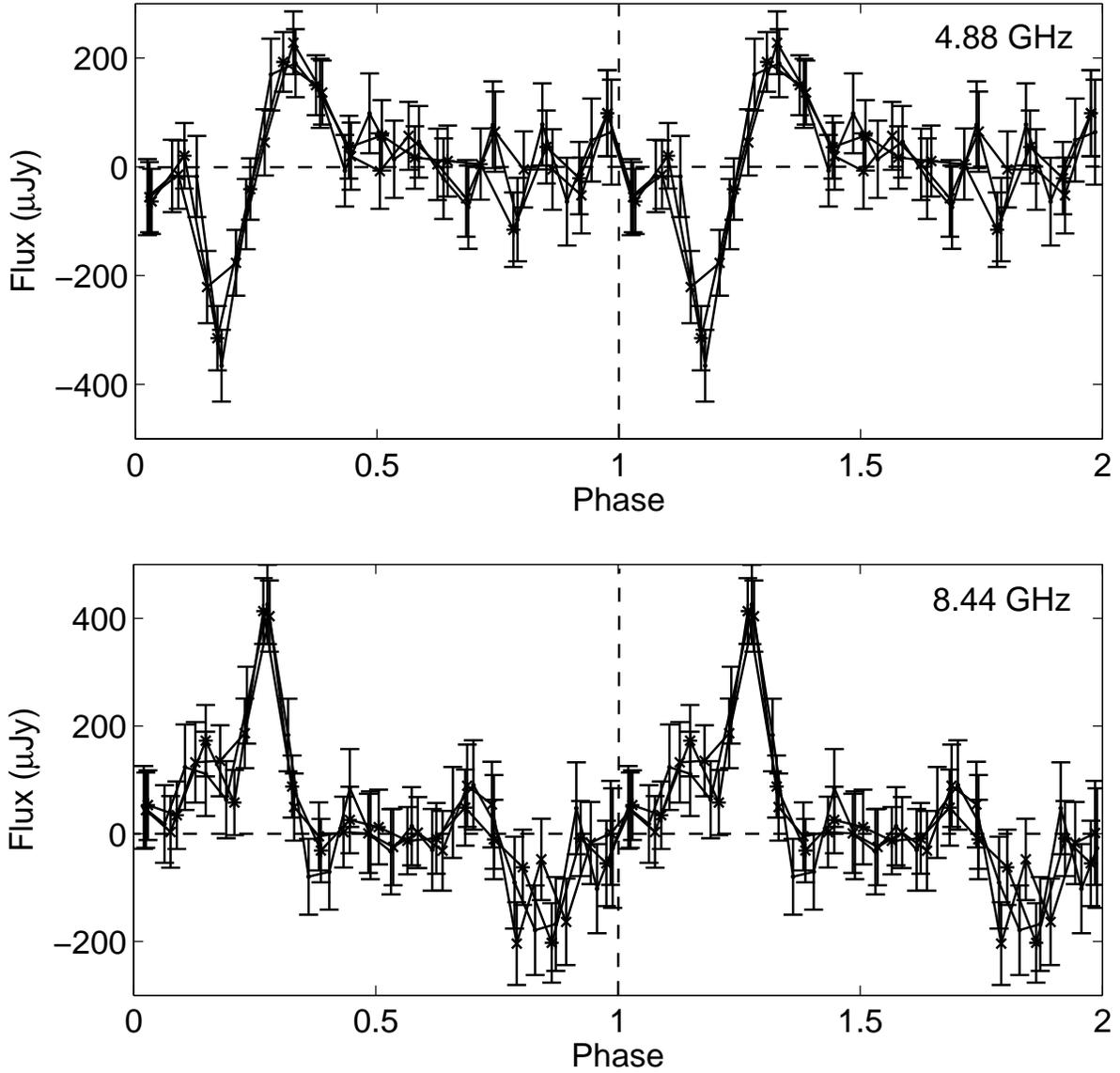}
\caption{Epoch folded light curves derived for the 4.88 GHz (\textit{top}) and 8.44 GHz (\textit{bottom}) Stokes V radio data from TVLM 513 with higher time resolutions of 6, 7 \& 8 minutes.  Two periods of the same epoch folded light curve are shown for clarity.  The large periodic peak in left circular polarization observed in the 4.88 GHz data has no corresponding counterpart in the 8.44 GHz data.  In fact, the emission is mildly right circular polarized suggesting a reversal in polarization with frequency.}     
\end{figure}




\end{document}